\documentstyle[12pt]{article}
\begin{document}
\newcommand{\pl}[1]{Phys.\ Lett.\ {\bf #1}\ }
\newcommand{\npb}[1]{Nucl.\ Phys.\ {\bf B#1}\ }
\newcommand{\prd}[1]{Phys.\ Rev.\ {\bf D#1}\ }
\newcommand{\prl}[1]{Phys.\ Rev.\ Lett.\ {\bf #1}\ }
\newcommand{\hepph}[1]{{\tt hep-ph/#1}}
\newcommand{\hepth}[1]{{\tt hep-th/#1}}

\newcommand{\drawsquare}[2]{\hbox{%
\rule{#2pt}{#1pt}\hskip-#2pt
\rule{#1pt}{#2pt}\hskip-#1pt
\rule[#1pt]{#1pt}{#2pt}}\rule[#1pt]{#2pt}{#2pt}\hskip-#2pt
\rule{#2pt}{#1pt}}

\newcommand{\Yfund}{\raisebox{-.5pt}{\drawsquare{6.5}{0.4}}}
\newcommand{\Ysymm}{\raisebox{-.5pt}{\drawsquare{6.5}{0.4}}\hskip-0.4pt%
        \raisebox{-.5pt}{\drawsquare{6.5}{0.4}}}
\newcommand{\Yasymm}{\raisebox{-3.5pt}{\drawsquare{6.5}{0.4}}\hskip-6.9pt%
        \raisebox{3pt}{\drawsquare{6.5}{0.4}}}

\begin{titlepage}
\begin{center}
{\hbox to\hsize{hep-th/9607021 \hfill  MIT-CTP-2543}

\bigskip

\bigskip

{\Large \bf  Supersymmetry Breaking  Through Confining
and Dual Theory Gauge Dynamics 
\footnotemark[1]      } \\

\bigskip

\bigskip

{\bf Csaba Cs\'aki, Lisa Randall,\footnotemark[2]  Witold Skiba}\\

\smallskip

{ \small \it Center for Theoretical Physics

Laboratory for Nuclear Science and Department of Physics

Massachusetts Institute of Technology

Cambridge, MA 02139, USA }

\bigskip
 
and\\
\smallskip
{\bf Robert G. Leigh}\\
\smallskip
{ \small \it Department of Physics

Rutgers University

Piscataway, NJ 08855-0849, USA}

 }

\vspace{1cm}
{\bf Abstract}\\
\end{center}
We show that theories in the confining, free magnetic, and conformal
phases can break supersymmetry through dynamical effects. To illustrate
this, we present theories based on the gauge groups 
$SU(n)\times SU(4)\times U(1)$ and $SU(n) \times SU(5) \times U(1)$
with the field content obtained by decomposing an $SU(m)$ theory with 
an antisymmetric tensor and $m-4$ antifundamentals.

\bigskip

\footnotetext[1]{Supported in part by DOE under cooperative 
                 agreement \#DE-FC02-94ER40818
                 and \#DE-FG05-90ER40599.}
\footnotetext[2]{NSF Young Investigator Award, Alfred P.~Sloan
Foundation Fellowship, DOE Outstanding Junior Investigator Award. }

\end{titlepage}

Recently, there has been a dramatic increase in the number of dynamical
supersymmetry breaking models~\cite{dnns,Poppitz,it,iss,us}. In part
this increase has been due to new methods for analyzing supersymmetric
theories~\cite{Seiberg,duality}. While many of the first models for
breaking supersymmetry had instanton or gaugino generated terms which
kept fields away from the origin~\cite{dnns,Poppitz,early}, recent work
has argued that  models in other phases can also break supersymmetry. In
 Ref.~\cite{iss}, it was argued that supersymmetry can be broken due to
confinement.  A nontrivial modification of the K\"ahler potential near
the origin removes the supersymmetry preserving minimum.  Alternatively,
models with a quantum modified moduli space can also break
supersymmetry~\cite{it}  because the supersymmetry preserving origin is
removed by the quantum modified constraint. Models in the conformal or
free magnetic phase can also break supersymmetry. In the models which
have been studied to now, these models broke supersymmetry through  an
O'Raifeartaigh  mechanism in the dual theory~\cite{Pouliot,PS} or strong
dynamics in the electric theory. A class of models described below is
distinguished by the fact that the  dynamics can be understood only in
the dual description where dynamical  effects are responsible for
supersymmetry breaking.

In a recent paper~\cite{us}, a new class of models was
studied which were based on a product group in which supersymmetry is
broken dynamically. There it was argued that supersymmetry breaking
could be understood as a collusion between separate dynamical effects
from the two nonabelian gauge groups. In the first example, the 4-3-1
model based on the gauge group $SU(4)\times SU(3)\times U(1)$, the exact
superpotential could be found and the model was an O'Raifeartaigh model
with both groups contributing to the final form of the superpotential.
In all cases, supersymmetry breaking could be understood by taking a
limit in which the gauge coupling of a confining gauge group is the
biggest coupling. In this limit, Yukawa couplings which were necessary
to lift flat directions turn into mass terms. Many flavors can be
integrated out and the gauge dynamics of the second nonabelian gauge
factor generated a superpotential which drives fields from the origin
leading to the breaking of supersymmetry.

In the particular models considered in Ref.~\cite{us}, other mechanisms
of supersymmetry breaking could appear as well in the limit that one of
the gauge couplings dominated. For example, in the particular case of
the 4-3-1 model supersymmetry breaking occurs in the strong $\Lambda_3$
limit through confinement, analogous to the mechanism of
Ref.~\cite{iss}. On the other hand, if some of the tree level terms are
removed, supersymmetry breaking appears due to a quantum modified
constraint~\cite{it}. Because of these additional descriptions, it was
not clear that the quantum modified constraint was not essential to
supersymmetry breaking.

In this paper, we show that analogous models in which each of the two
groups is in one of a confining, free magnetic, or conformal phase (in
the limit that we neglect the other coupling) also break supersymmetry,
through a conspiracy of dynamical effects from the two gauge groups.
Naively, it would appear that such models should allow fields to go to
the origin. However, because of the tree-level superpotential and
dynamics of one group, the other group can generate a dynamical
superpotential in the infrared which forbids the origin and yields
supersymmetry breaking.

It is interesting that models in which the theory must be analyzed at
low energy in the dual phase can break supersymmetry. It is not
essential for the number of flavors to be so small that a dynamical
superpotential, a quantum modified constraint, or even confinement
occurs in the electric theory. This suggests the possibility of a much
larger class of supersymmetry breaking models because of the much less
restrictive condition on the  size of the initial particle content.

The two models we present in this paper are obvious generalizations of
the models considered in Ref.~\cite{us}. Analogously to the n-3-1
models, supersymmetry breaking can be understood as a result of Yukawa
couplings and strong dynamics which  make flavors of the second gauge
group heavy. In the resulting theory, the origin is forbidden because of
a dynamical superpotential from the second gauge group. The mechanism is
in some sense independent of the number of flavors in the initial
theory. We present two classes of models to illustrate this. In the
first class of models, in which one of the gauge groups is confining,
supersymmetry breaking occurs through a conspiracy of gauge effects. We
then consider a model which must be analyzed in the dual phase. The
supersymmetry breaking dynamics for this model is remarkably similar to
that of the confining theory, as we will show below.
  
The fields of the first model can be obtained by decomposing $SU(n+4)$
model with an antisymmetric tensor~\cite{early} into its $SU(n)\times
SU(4)\times U(1)$ subgroup. The field content is
\begin{eqnarray}
\Yasymm & \rightarrow & A(\Yasymm,1)_8 + 
                       a(1,\Yasymm)_{-2 n} +
                       T(\Yfund,\Yfund )_{4-n} \nonumber \\
n \cdot \overline{\Yfund} & \rightarrow & \bar{F}_{I} 
       (\overline{\Yfund},1)_{-4} + \bar{Q}_{i}
(1,\overline{\Yfund})_n,
\end{eqnarray}
where $i,I=1,\ldots,n$. We take the tree-level superpotential to be
\begin{eqnarray}
\label{wtree}
W_{tree}=A\bar{F}_1\bar{F}_2+A\bar{F}_3\bar{F}_4+\ldots
+A\bar{F}_{n-2}\bar{F}_{n-1}
\;\;\;\;\;\;\;\;\;\;\;\nonumber \\
+a\bar{Q}_2\bar{Q}_3+a\bar{Q}_4\bar{Q}_5+\ldots
+a\bar{Q}_{n-1}\bar{Q}_1 +
T\bar{F}_1\bar{Q}_1+\ldots +T\bar{F}_n\bar{Q}_n.
\end{eqnarray}
A detailed analysis along the lines of Ref. \cite{us} shows that this
superpotential lifts all flat directions. The relative shift of the
indices in the $A\bar{F}\bar{F}$ and $a\bar{Q}\bar{Q}$ terms is
important. Without this shift not all flat  directions  are lifted. This
superpotential preserves an $R$-symmetry which is anomalous only under
the $U(1)$ gauge group.

We analyze this theory in the limit where $\Lambda_n\gg\Lambda_4$. The
$SU(n)$ field content is an antisymmetric tensor, four fundamentals and
$n$ antifundamentals which give confining gauge dynamics. Below
$\Lambda_n$, the effective degrees of freedom are the $SU(n)$
invariants~\cite{Pouliot}
\begin{eqnarray}
\label{suninvariants}
&& X_{IJ}=A^{\alpha \beta}\bar{F}_{\alpha I}\bar{F}_{\beta J}
\nonumber \\
&& \bar{B}=\bar{F}_{\alpha_1 1}\ldots \bar{F}_{\alpha_n n}
\epsilon^{\alpha_1 \ldots \alpha_n} \nonumber \\
&& (B_1)^a=T^{\alpha_1 a} A^{\alpha_2 \alpha_3}\ldots 
A^{\alpha_{n-1} \alpha_n} \epsilon_{\alpha_1 \ldots \alpha_n}
\nonumber \\
&& (B_3)_a=\epsilon_{abcd} T^{\alpha_1 b}T^{\alpha_2 c}T^{\alpha_3 d}
 A^{\alpha_4 \alpha_5}\ldots 
A^{\alpha_{n-1} \alpha_n} \epsilon_{\alpha_1 \ldots \alpha_n}
\nonumber \\
&& M_I^a=T^{\alpha a}\bar{F}_{\alpha I},
\end{eqnarray}
plus the $SU(n)$ singlets $a$ and $\bar{Q}_i$.

The superpotential is the sum of  the tree-level terms from
Eq.~(\ref{wtree}) and the confining superpotential~\cite{Pouliot}.
\begin{eqnarray}
\label{sun}
W&=&X_{12}+\ldots +X_{n-2,n-1}+
a\bar{Q}_2\bar{Q}_3+\ldots +a\bar{Q}_{n-1}\bar{Q}_1 +\nonumber \\
&&M_1\bar{Q}_1+\ldots +M_n\bar{Q}_n+ \frac{1}{\Lambda_n^{2n-1}}
\Big(B_{3 a} M_{I_1}^a X_{I_2I_3}\ldots  X_{I_{n-1}I_n}
\epsilon^{I_1\ldots I_n} \nonumber \\
&& + B_1^a M_{I_1}^b M_{I_2}^c M_{I_3}^d X_{I_4I_5}\ldots
X_{I_{n-1}I_n}
\epsilon^{I_1\ldots I_n}\epsilon_{abcd} +\bar{B}B_1^aB_{3 a} \Big),
\end{eqnarray}
where small Latin letters denote $SU(4)$ indices.

Note that in the confined theory, some of the Yukawa couplings  have
become mass terms. To deduce the infrared theory, we integrate out all
massive fields. It is technically difficult to integrate out the fields
using the full superpotential from Eq.~(\ref{sun}). For simplicity we
set the couplings of all $a\bar{Q}\bar{Q}$ terms to zero.  We will argue
based on symmetries that the models with the additional baryon operators
included still break supersymmetry. It should be noted that the flat
directions now present classically are lifted  in the quantum theory
\cite{it}, which is presumably a valid supersymmetry breaking model as
well.
 
Because we have integrated out  $n$ massive flavors, the  $SU(4)$ theory
at low energy has an antisymmetric tensor and  only one flavor. This
theory dynamically generates a  superpotential. The   low-energy
superpotential is therefore
\begin{eqnarray}
\label{effective}
W_{\rm eff}=X_{12}+\ldots
+X_{n-2,n-1}+\frac{1}{\Lambda_n^{2n-1}}\bar{B}m+
\left[\frac{\tilde{\Lambda}_4^5}{{\rm Pf}a \, m}\right]^{\frac{1}{2}},
\end{eqnarray}
where ${\rm Pf}a=a^{ab}a^{cd}\epsilon_{abcd}$, $m=B_1^aB_{3 a}$, and
$\tilde{\Lambda}_4$ is the dynamical scale of the effective one flavor
$SU(4)$ theory.  The equations of motion have set  most terms to zero in
the $\Lambda_n$ dependent term. The $\bar{B}$ equation of motion would
set $m=0$. However, this is inconsistent with  the
$\left[\frac{\tilde{\Lambda}_4^5}{{\rm Pf}a \, m}\right]^{\frac{1}{2}}$
term in the superpotential, which drives $m$ from the origin in a theory
with no flat directions. Therefore, we conclude that the equations of
motion are contradictory, and supersymmetry is dynamically broken.

We have argued that supersymmetry is broken in the theory with
$\gamma^{ij}=0$, where $\gamma^{ij}$ is the coefficient of the 
$a\bar{Q}\bar{Q}$ operators in the tree-level superpotential. It is
clear that even with nonzero $\gamma^{ij}$, supersymmetry is still
broken. From symmetries, it can be shown that the neglected terms can
correct the superpotential by a power series in
\begin{equation}\label{eq:correc}
{\cal A}=\Lambda_n^{-2n+1}({\rm Pf}a)^{\frac{1}{2}}
(X_{IJ})^{\frac{n-2}{2}}m^{\frac{1}{2}}
(\gamma^{ij})(m^{iI})^{-2},
\end{equation}
where $m^{iI}$ is the coefficient of the  $T\bar{F}_I\bar{Q}_i$
operators. For small $\gamma$, these terms could only  give a
sufficiently large contribution to cancel a nonzero $F$-term at field
values larger than $\Lambda_n$. In this case, the theory  should have
been analyzed in the Higgs phase, which is clearly inconsistent with
supersymmetry since there were no flat directions.

As an aside, we note that in the version of the theory without the
$a\bar{Q}\bar{Q}$ terms in the superpotential (and hence without the
corrections of Eq.~(\ref{eq:correc})), there is an additional source of
supersymmetry breaking. The  terms $X_{12}+\ldots +X_{n-2,n-1}$ in the
superpotential lead to supersymmetry breaking due to confinement, as
described in Ref.~\cite{iss}. Here we emphasize the first argument for
supersymmetry breaking, which generalizes beyond confining models, as we
describe below.

Next, we consider theories based on the gauge group $SU(n) \times SU(5)
\times U(1)$ ($n$ even) obtained by reducing the gauge group of the
$SU(n+5)$ theory with an antisymmetric tensor and $n+1$
antifundamentals. The mechanism of supersymmetry breaking will turn
out to be very similar to the previous models, despite the very
different gauge dynamics.

The field content is
\begin{eqnarray}
\Yasymm & \rightarrow & A(\Yasymm,1)_{10} + 
                       a(1,\Yasymm)_{-2n} +
                       T(\Yfund,\Yfund )_{5-n} \nonumber \\
(n+1)\cdot \, \overline{\Yfund} & \rightarrow & \bar{F}_{I} 
       (\overline{\Yfund},1)_{-5} + \bar{Q}_{i}
(1,\overline{\Yfund})_n,
\end{eqnarray}
where $i,I=1,\ldots,n+1$. The tree-level superpotential is
 \begin{eqnarray}
\label{tree651}
W_{tree}&=&A\bar{F}_1\bar{F}_2+
\ldots+A\bar{F}_{n-1}\bar{F}_{n}
+a\bar{Q}_2\bar{Q}_3+
\ldots+a\bar{Q}_{n}\bar{Q}_1 +\nonumber \\
& & T\bar{F}_1\bar{Q}_1+\ldots +T\bar{F}_{n+1}\bar{Q}_{n+1}.
\end{eqnarray}
Again a detailed analysis verifies the absence of flat directions.

The $SU(5)$ gauge group has an antisymmetric tensor and $n$ flavors
while the $SU(n)$ has an antisymmetric tensor and five flavors. The
$SU(5)$ group is in the conformal regime while the $SU(n)$ group is in
the free magnetic phase. Although it seems more obvious to dualize the
$SU(n)$ which is in the free magnetic phase it is simpler to dualize the
gauge group $SU(5)$, as it has an odd number of colors. This duality
will increase the number of $SU(n)$ flavors by $n-3$ which takes the
theory out of the free magnetic phase.

The dual description of $SU(5)$ with an antisymmetric tensor and $n$
flavors is an $SU(n-3)\times Sp(2n-8)$ gauge theory\cite{Pouliot} with
the field content given in Table~\ref{table}.

\begin{table}
\begin{equation}
\begin{array}{c|cccccc}
  & SU(n-3) & Sp(2n-8) & SU(n)   & U(1) & SU(n+1)_{\bar{Q}} &
SU(n+1)_{\bar{F}} \\ \hline
A &  1    &    1  & \Yasymm & 10  & 1 & 1\\
\bar{F} & 1 & 1 & \overline{\Yfund} & -5 & 1 & \Yfund \\
x & \Yfund & \Yfund & 1 & 0 & 1 & 1 \\
p & \Yfund & 1 &  1 & 5n & 1 & 1\\
\bar{a} & \overline{\Yasymm} & 1 & 1 & 0 & 1 & 1 \\
\bar{q} & \overline{\Yfund} & 1 & \overline{\Yfund} & -5 & 1 & 1 \\
l & 1 & \Yfund & 1 & 0 & \overline{\Yfund} & 1\\
M & 1 & 1 & \Yfund & 5 & \Yfund & 1 \\
H & 1 & 1 & 1 & 0 & \Yasymm & 1\\
B_1 & 1 & 1 & \Yfund & 5(1-n) & 1 & 1\\
\end{array}
\end{equation} 
\caption{The field content of the $SU(n)\times SU(5)\times U(1)$ theory
after dualizing the $SU(5)$ gauge group. \label{table}} 
\end{table} 

The $SU(n-3)\times Sp(2n-8)$ gauge group in Table~\ref{table} is the dual
of the $SU(5)$ gauge group, while the $SU(n)\times U(1)$ is the
remaining original gauge group unchanged by the duality transformation.
The $SU(n+1)_{\bar{Q}} \times SU(n+1)_{\bar{F}}$ global symmetries are
the non-abelian global symmetries of the original $SU(n)\times
SU(5)\times U(1)$ theory.

The superpotential consists of the terms corresponding to the
tree-level superpotential of Eq.~(\ref{tree651}) and the terms arising
from the duality transformation. It is given by
\begin{eqnarray}
W&=& A\bar{F}_1\bar{F}_2+
\ldots +A\bar{F}_{n-1}\bar{F}_{n} +
H_{23}+\ldots +H_{n 1} + M_1 \bar{F}_1 + \ldots + \nonumber \\
&& M_{n+1} \bar{F}_{n+1}
+ M \bar{q} l x + H l^2 + B_1 p \bar{q} + \bar{a} x^2.
\end{eqnarray}


As in the $SU(n)\times SU(4)\times U(1)$ models, some of the tree-level
Yukawa terms are mapped into mass terms in the dual description. To
simplify the theory we again set the coefficients of the 
$A\bar{F}\bar{F}$ operators to zero, though
in this case it is not difficult to leave them in.
 With this simplification, one can easily integrate out 
the massive flavors of $SU(n)$ since the $\bar{F}_I$ equations of motion 
set all $M$'s to zero. There is just one $SU(n)$ flavor remaining and thus
there is a dynamically generated term in the superpotential from the
$SU(n)$ dynamics. The effective low-energy superpotential is
\begin{equation}
\label{weff}
W=H_{23}+ H_{45}+ \ldots +H_{n1}+ H l^2 + \bar{a} x^2 + \tilde{M} p +
    \frac{\tilde{\Lambda}_n^{n+1}}{(\tilde{M} \tilde{X}^{(n-4)/2} {\rm
Pf}\,A)^{1/2}},
\end{equation}
where $\tilde{M}=B_1 \bar{q}$, $\tilde{X}= A\bar{q}\bar{q}$ and ${\rm
Pf}\,A=A^{n/2}$, while $\tilde{\Lambda}_n$ is the effective $SU(n)$
scale. This superpotential looks very much like the one in
Eq.~(\ref{effective}), with $\tilde{M}$ playing the role of $m$ and p
the role of $\bar{B}$. The equations of motion are again contradictory. 
We again conclude that supersymmetry is broken.

The above analysis neglected the $Sp(2n-8)$ group that appears from
dualizing the $SU(5)$ group. This group is however Higgsed by the VEV's
of the $l$ fields as a result of the $H$ equations of motion and the
terms linear in $H$ in the superpotential. Although instanton terms can
be generated in the broken $Sp(2n-8)$ group, these will not involve the
fields $\tilde{M},\tilde{X},{\rm Pf}A$ or $p$ and therefore do not
affect the proof of dynamical supersymmetry breaking given above. The
$Sp(2n-8)$ dynamics seems to be irrelevant to the analysis of the model.

The dynamics of the general $SU(n)\times SU(m)\times U(1)$ models 
($n,m\geq 5$) obtained in the same way is very similar to that of the
$SU(n)\times SU(5)\times U(1)$ model, if one dualizes the $SU(n)$
corresponding to odd $n$. We expect that a similarly constructed
tree-level superpotential lifts all flat directions. One can then show
that the resulting low-energy superpotential is in one-to-one
correspondence to the superpotential of Eq.~(\ref{weff}), with the
remaining gauge group being $SU(m-3)\times Sp(2m-8)\times SU(m)\times
U(1)$ ($m$ is even), which is obtained by dualizing the original $SU(n)$
group. Since the superpotential is exactly of the same form as the one
in Eq.~(\ref{weff}) we conclude that the general $SU(n)\times
SU(m)\times U(1)$ models break supersymmetry as well.

The similarities between the  $SU(n)\times SU(4)\times U(1)$ and
$SU(n)\times SU(5)\times U(1)$ models  is intriguing. In both models,
the dynamics of the $SU(n)$ group leads to additional flavors of the
second gauge group, in one case due to confinement, and in the other
case, due to the dual description. In both cases, some of the tree level
terms are mapped into mass terms due to dynamical effects in the $SU(n)$
gauge group. After integrating out these massive flavors the other gauge
group has only a single flavor remaining besides the antisymmetric
tensor and produces a dynamically generated superpotential. This
dynamical superpotential together with a piece of the superpotential
from the strong dynamics of the first group breaks supersymmetry. Thus
supersymmetry breaking in these theories involves a subtle interplay
between the gauge dynamics of both groups and the tree-level
superpotential.

That these theories (and presumably the general $SU(n)\times SU(m)\times
U(1)$ models as well) break supersymmetry suggests the existence of
still more models of dynamical supersymmetry breaking. The flavor
content of these models can be much larger than one would naively have
anticipated by the requirement of a dynamical superpotential, because
Yukawa couplings or other interactions in the presence of strong
dynamics can change the phase of the theory in the infrared. The
low-energy description might then have sufficiently few flavors to break
supersymmetry dynamically. 
 
\medskip
\noindent {\bf Acknowlegements:}  Conversations with Erich Poppitz and
Riccardo Rattazzi are gratefully acknowledged. We note that, after this
work was completed, a similar discussion of supersymmetry breaking 
appeared in Ref. \cite{PST} in the context of product group theories
with fundamental matter.

\end{document}